# Efficient Switching of 3-Terminal Magnetic Tunnel Junctions by the Giant Spin Hall Effect of Pt$_{85}$Hf$_{15}$ Alloy

Minh-Hai Nguyen[1], Shengjie Shi[1], Graham E. Rowlands[2], Sriharsha V. Aradhya[1*],

Colin L. Jermain[1], D. C. Ralph[1,3], R. A. Buhrman[1†]

[1] *Cornell University, Ithaca, New York 14853, USA*

[2] *Raytheon BBN Technologies, Cambridge, Massachusetts 02138, USA*

[3] *Kavli Institute at Cornell, Ithaca, New York 14853, USA*

## ABSTRACT

Recent research has indicated that introducing impurities that increase the resistivity of Pt can enhance the efficiency of the spin Hall torque it generates. Here we directly demonstrate the usefulness of this strategy by fabricating prototype 3-terminal in-plane-magnetized magnetic tunnel junctions that utilize the spin Hall torque from a Pt$_{85}$Hf$_{15}$ alloy, and measuring the critical currents for switching. We find that Pt$_{85}$Hf$_{15}$ reduces the switching current densities compared to pure Pt by approximately a factor of 2 for both quasi-static ramped current biases and nanosecond-scale current pulses, thereby proving the feasibility of this approach to assist in the development of efficient embedded magnetic memory technologies.

[*] Currently at Western Digital Corp., Fremont, CA 94539
[†] Email address: buhrman@cornell.edu

Magneto-resistive random access memories (MRAM) have drawn a great deal of attention from the electronics and spintronics communities due to their inherent nonvolatility which enables long data retention and low standby power[1]. Currently the prevalent type of MRAM device is the two-terminal magnetic tunnel junction (2T-MTJ) in which a partially spin-polarized electrical current is applied through the MTJ layers to switch the magnetization of the ferromagnetic (FM) free layer parallel (P) or anti-parallel (AP) to the fixed magnetization of the FM reference layer via spin transfer torque[2]. Recently three-terminal MTJ (3T-MTJ) structures based on the spin Hall effect (SHE)[3–5] have been proposed[6] as a means of achieving faster write and read times while avoiding the issues of wear-out and/or breakdown of the tunnel barrier due to the write current and of read disturb errors associated with the two-terminal geometry. The separation of the write and read paths furthermore allows additional flexibility in optimizing the performance of the devices without the trade-offs required in their two-terminal counterparts, although they are more limited in scaling to ultra-high density. 3T-MTJs can have the free layer magnetization either perpendicular to the plane or in the sample plane, with different modes of spin-Hall-driven reversal. The former, which switch in response to the effective spin-torque field, can be reversed quite quickly but this requires quite high current densities[7,8]. In addition, fast reversal that is highly reliable has yet to be demonstrated with such devices. Here we will consider devices with in-plane magnetic anisotropy that reverse due to anti-damping spin torque and hence have much lower switching current densities. Spin Hall switching of in-plane-magnetized 3T-MTJs was initially expected to be relatively slow and unreliable for short pulses based on experience with in-plane magnetized 2T-MJT devices[9]. However it has been recently demonstrated[10] that spin Hall switching of 3T-MTJs can be both surprisingly fast (2 ns) and reliable (write error rates < $10^{-5}$). This performance can be understood by accounting for the role of the field-like torque and micromagnetic dynamics in reducing the incubation delay[11]. This excellent performance suggests that the 3T-MTJs could be an attractive candidate for fast, embedded memory applications where ultra-high density is not essential, provided a sufficiently low pulsed write current can be obtained.

The critical current density for anti-damping spin torque switching of 3T-MTJ devices with in-plane anisotropy is directly proportional to the FL's effective Gilbert magnetic damping $\alpha$, which is the sum of the intrinsic damping of the FL and any enhanced damping due to non-ideal effects that cause interfacial "spin memory loss" (SML), and/or to spin pumping into the adjacent heavy metal layer. The critical current density is also inversely proportional to the damping-like spin torque efficiency $\xi_{DL} = T_{int}\theta_{SH}$, where $T_{int} < 1$ is the spin transparency of the interface and $\theta_{SH}$ is the spin Hall ratio of the channel. It has been reported that when SML is a significant factor, due to the interfacial interaction of the HM with the FM, the resulting increase in the damping $\alpha$ can be reduced by passivating the interface with atomic insertion layers of another metal,[12,13] without overly reducing the net interfacial transparency $T_{int}$[14,15]. Extensive efforts have also been made to improve the efficiency of spin-Hall switching by identifying and utilizing HM thin film materials with large $\theta_{SH}$, such as Pt[12,16], $\beta$-Ta[6], $\beta$-W[17], Pd[18], their oxides[19,20] and binary alloys[21–27]. Recent research has shown both that the intrinsic SHE is the dominant mechanism in Pt, with the consequence that the spin Hall ratio scales linearly with the Pt resistivity $\rho_{Pt}$, or $\theta_{SH} = \rho_{Pt}\sigma_{SH}$ where $\sigma_{SH}$ is the spin Hall conductivity, and that Pt has the highest $\sigma_{SH}$ of the known conventional metals[28–32]. This opens up the opportunity to further enhance $\theta_{SH}$ of Pt by alloying to increase its resistivity. Previously we have demonstrated through spin-torque measurements on perpendicularly magnetized samples[21] that the Pt$_{85}$Hf$_{15}$ alloy exhibits 2-3 times higher spin torque efficiencies, depending on the particular HM/FM interface, than pure Pt. Here we investigate directly whether $Pt_{85}Hf_{15}$ can indeed improve the performance of prototype 3T-MTJs compared to pure Pt, by measuring the fast-pulse switching characteristics of in-plane 3T-MTJ devices that utilize a $Pt_{85}Hf_{15}$ channel, and with and without a Hf insertion layer.

We report switching studies on two series of in-plane 3T-MTJ devices, consisting of multilayer stacks of substrate | Ta(1) | PtHf(6) | Hf(0 or 0.7) | FeCoB(1.4) | MgO | FeCoB(1.2) | Ta(0.2) | FeCoB(1.2) | FeCo(1) | Ru(0.85) | FeCo(2.5) | IrMn(7) | Ru(4) (thicknesses in nm, PtHf = $Pt_{85}Hf_{15}$, FeCoB = $Fe_{60}Co_{20}B_{20}$), that were DC and RF (for the MgO) magnetron sputter-deposited onto thermally oxidized high-resistivity Si wafers by Canon ANELVA, Inc. The 1 nm Ta layer at the bottom provides a smoothing nucleation layer for the growth of the 6 nm PtHf channel layer. The magnetization of the FeCoB FM reference layer is pinned by the IrMn layer atop the FM/Ru/FM synthetic antiferromagnetic structure. The stacks are patterned into 3T-MTJ devices and annealed at 360 C in an applied magnetic field by the same processes previously reported[10]. Fig. 1(a) shows a tilted electron microscopy image of the channel and MTJ pillar (with e-beam resist on top) of a device after the ion-mill etching of the pillar but before deposition of a field insulator and top contacts. The size of the channel was measured by atomic force microscopy to be 440 nm × 600 nm. The channel resistance was 2.5 kΩ, more than 2 times higher than similar devices having a pure Pt channel[10] due to the higher resistivity of the PtHf channel;[21] $\rho_{PtHf}$(6 nm) ≈ 110 μΩ · cm, $\rho_{Pt}$(5 nm) ≈ 45 μΩ · cm. The design size of the MTJ pillars was 45 nm × 190 nm (aspect ratio = 1:4.2).

The circuit for our initial characterization of devices using direct-current (DC) measurements is illustrated in Fig. 1(a). All measurements reported here were performed at room temperature. The state of the MTJ was determined by reading the voltage drop across the MTJ by a lock-in amplifier using a small oscillating current (≤ 1 µA). Fig. 1(b) shows the magnetic hysteresis (minor) loops of the MTJ when an external field is swept parallel to the easy axis of the MTJ by a Helmholtz pair electromagnet. All devices exhibit sharp field switching behavior. The centers of the loops are located at $H_{off}$ = 0 mT and -4.5 mT, respectively, for samples without and with the Hf spacer. The coercive field of the devices without the spacer was ~ 8.5 mT, while that of the ones with the spacer were lower, ~3.2 mT.

To estimate $I_0$, the critical current in the absence thermal fluctuations, and the thermal stability factor, we used a fixed field $H_{off}$ to bias the free layer into the midpoint of the bi-stable state and applied a ramped DC current to the channel to manipulate the magnetization of the free layer. An example of the measured MTJ resistance versus the DC current is shown in Fig. 1(c). Note that in our convention, a positive external field is in the same direction as the current-induced Oersted field acting on the free layer (and therefore also the same as the moment direction of the incident spin current). Due to thermal fluctuations the measured DC critical current $I_c$ depends on the current ramp rate, with the expected behavior being given by the standard macrospin model as [33,34]:

$$I_c = I_0 \left[ 1 + \frac{1}{\Delta} \ln\left( \left|\frac{\dot{I}}{I_0}\right| \tau_0 \Delta \right) \right], \quad (1)$$

where $\Delta = E / k_B T$ is the thermal stability factor (E: barrier energy, $k_B$: Boltzmann constant, T: temperature = 300 K), $\dot{I}$ the current ramp rate, and $\tau_0$ the thermal fluctuation time (taken to be 1 ns). The current ramp-rate results and fits to equation (1) are shown in Fig. 1(d). The fit values averaged over two identical samples are $I_0$ = ±0.56 ± 0.03 mA and $\Delta$ = 70 ± 13 for the device without the spacer and $I_0$ = ±0.38 ± 0.03 mA, $\Delta$ = 36 ± 8 with the Hf spacer (positive currents drive the P-to-AP transition and negative currents AP-to-P). These results, together with the values of magnetic damping $\alpha$ and effective field $M_{eff}$ estimated by ferromagnetic resonance measurements (see section S1 of Supplemental Information - SI[35]), are listed in Table 1 for our two types of PtHf devices, with a comparison to the sample reported in Ref. [10] having a 5 nm pure Pt channel, the same MTJ aspect ratio, and a slightly thicker free layer (1.6 nm). The spin torque efficiency $\xi_{DL}$ is calculated from the above parameters by the relation[36]:

$$\xi_{DL} = \frac{2e}{\hbar} \mu_0 M_s t_{FeCoB} \alpha \left( H_c + M_{eff}/2 \right) / J_0^{DC}, \qquad (2)$$

where $\mu_0 = 4\pi \times 10^{-7}$ N/A$^2$ is the vacuum permeability, $M_s$ the saturation magnetization, assumed to be $1.2 \times 10^6$ A/m (*i.e.*, no magnetic dead layer)[10] for all samples, $H_c$ the coercive field determined from the minor loops shown in Fig. 1(b) and $J_0^{DC}$ the absolute value of the DC critical current density in the absence thermal fluctuations.

From Table 1, it is readily seen that our PtHf alloy samples with a 1.4 nm FL have a consistently lower $M_{eff}$ than the Pt sample with the 1.6 nm FL, which is attributable to the stronger impact of the interfacial perpendicular magnetic anisotropy energy at the FL/MgO interface over the bulk in-plane anisotropy of the thinner FL. The magnetic damping constants of the devices having the Hf spacer are about 30% smaller than that of the one without the spacer which we attribute to the role of the Hf spacer in reducing the interfacial spin memory loss [12,37]. Most importantly, the spin torque efficiencies of the devices having the PtHf alloy channel are ≈ 0.10, about 2 times higher than that of the one having the Pt channel, which is consistent with the results measured on perpendicularly magnetized Pt/Co bilayers[21]. Finally, in comparison to the sample with the Pt channel and Hf spacer, these higher values of spin torque efficiency in combination with lower $M_{eff}$ and slightly thinner free layer result in ~2 times lower $J_0^{DC}$ for the PtHf device without the spacer, and ~3 times lower for the device with the Hf spacer, despite the spin attenuation of the Hf. We also note that the thermal stability factor, as well as the coercive field, of the PtHf sample having a Hf spacer is nearly 2 times lower than that of the one without a spacer. Since the structural difference between the two series of samples is the presence of a Hf spacer, we speculate that the difference in the thermal stability and coercive field is due to some as yet undetermined effect of the Hf layer on the micromagnetic details of the FM free layer.

As a more direct test of the performance for magnetic memory applications, we performed nanosecond-scale pulsed switching measurements on the PtHf devices. The measuring circuit and method are described in detail in sections S2 and S3 of SI[35]. Measurements were performed under a fixed applied magnetic field $H_{\text{off}}$ to position each device at the center of its minor loop. Figure 2(a) shows the resulting phase diagrams for both AP-to-P and P-to-AP switching for a device with the Hf spacer, for varying pulse duration and current amplitude (calculated from the voltage amplitude of the pulse and the channel resistance and geometry, taking into account the pulse reflection due to the impedance mismatch between the the 50 Ω probe and the channel). Here each switching probability is averaged over 1000 switching attempts. The 50% switching probability points are plotted in Fig. 2(b) for PtHf devices both with and without the Hf spacer.

As discussed in the previous reports[10,11] of fast switching in in-plane magnetized 3T-MTJ devices the FL reversal of these PtHf devices can be achieved using shorter pulses than expected from a rigid domain, macrospin model[36,38] for anti-damping excitation. Nevertheless, the predictions of that model can be useful in characterizing the short-pulse behavior at least for comparative purposes. By fitting the 50% probability points to the macrospin model:

$$J_{50\%} = J_0^{\text{pulse}}(1 + t_{50\%} / t_0), \quad (3)$$

we estimated the values of (pulsed) critical current density $J_0^{\text{pulse}}$ and time $t_0$, as listed in Table 1. The deviation of the fitted lines shown in Fig. 2(b) from the data points is another indication that the switching mechanism is not fully consistent with anti-damping excitation of a rigid, single domain, but, as indicated by micromagnetic modeling[11], the result of the rapid evolution of magnetic sub-domains within the free layer as driven by anti-damping spin torque and the field-like torques induced by the electrical and/or spin current. With this caveat, as listed in Table 1, the PtHf device without spacer exhibits 1.4 times lower $J_0^{\text{pulse}}$ and the one with the Hf spacer nearly 2 times lower than that of the one with pure Pt channel, qualitatively consistent with the corresponding critical currents for quasi-static switching.

The short-pulse behavior of PtHf devices in fact indicates an even faster characteristic switching speed in comparison to the Pt counterpart. This is quite encouraging since the micromagnetic analysis of the short-pulse switching of Pt 3T-MTJ devices indicated that the in-plane field generated by the pulse current in the spin Hall channel was likely an important factor in speeding the non-uniform magnetic reversal process. Given the enhanced spin torque efficiency for the PtHf devices one might have assumed that the lower pulse current that is required here would make this enhancement less effective, but that does not appear to be the case. The fast characteristic switching time is of course a very positive finding for application and motivates further study on the micromagnetic switching mechanism in 3T-MTJ structures.

In conclusion, we have demonstrated an enhanced spin torque efficiency of $Pt_{85}Hf_{15}$ alloy films, arising from the intrinsic SHE in Pt-based material, for nanoseond pulsed-current switching of 3T-MTJ devices having PtHf channels. The efficiency is further enhanced with the insertion of a thin (0.7 nm) Hf spacer to reduce interfacial magnetic damping and interfacial spin memory loss. We achieve 2 times lower critical current density for nanosecond scale pulse current switching compared to that of a Pt channel, and confirm a 2 times higher spin torque efficiency. These results suggest the opportunity to further improve the efficiency of Pt-based spin-orbit torque magnetic devices by alloying and additional interface engineering[39].

The research is based upon work supported by the Office of the Director of National Intelligence (ODNI), Intelligence Advanced Research Projects Activity (IARPA), via contract W911NF-14-C0089. This work was also supported in part by the Cornell Center for Materials Research with funding from the NSF MRSEC program (DMR-1719875). This work was performed in part at the Cornell NanoScale Facility, a member of the National Nanotechnology Coordinated Infrastructure (NNCI), which is supported by the National Science Foundation (Grant ECCS-1542081).

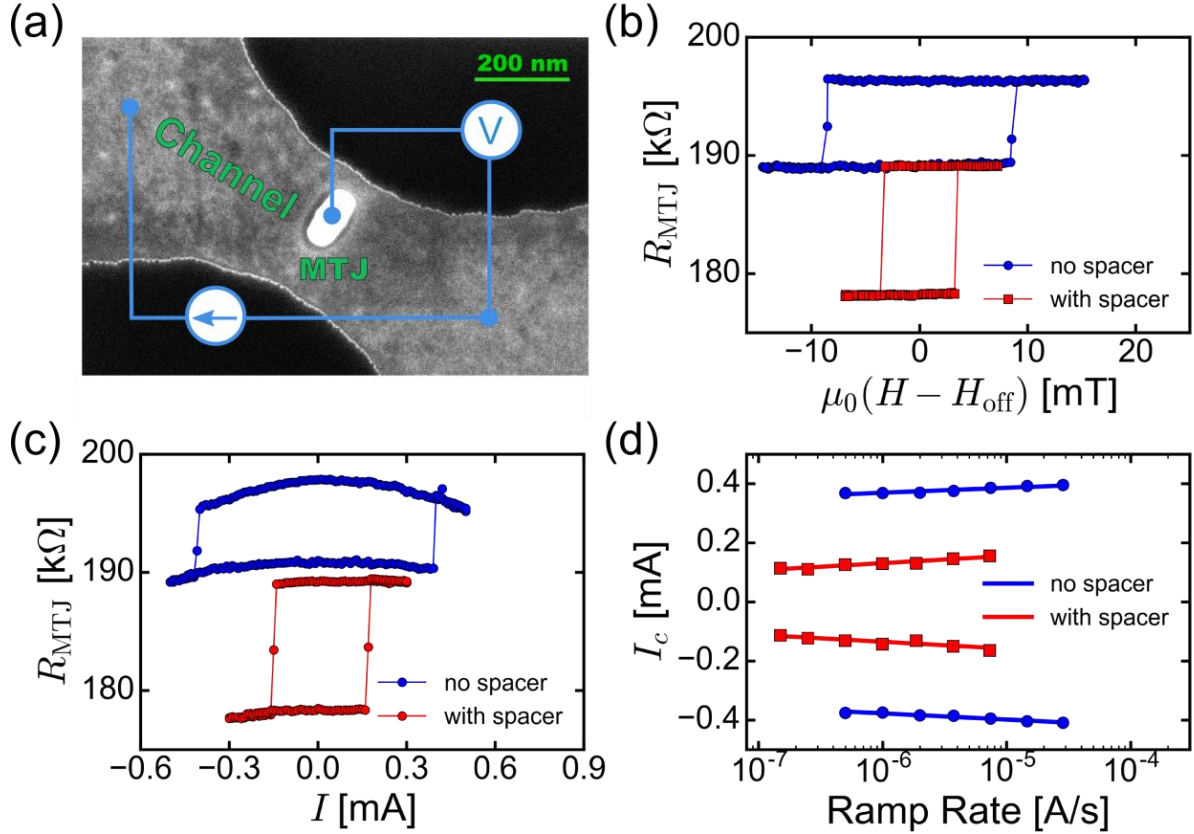

**Figure 1:** (a) Tilted electron microscopy image of a 3T-MTJ device, and the basic schematic of the circuit for DC measurements. (b) Magnetic minor loop and (c) DC current switching behavior of the device without the Hf spacer (blue) and with a 0.7 nm Hf spacer (red). The lines are guide to the eyes. (d) DC critical currents versus current ramp rates for P-to-AP (positive currents) and AP-to-P (negative currents) switching. The solid lines show fits to the thermally activated switching model. The fit parameters are summarized in Table 1.

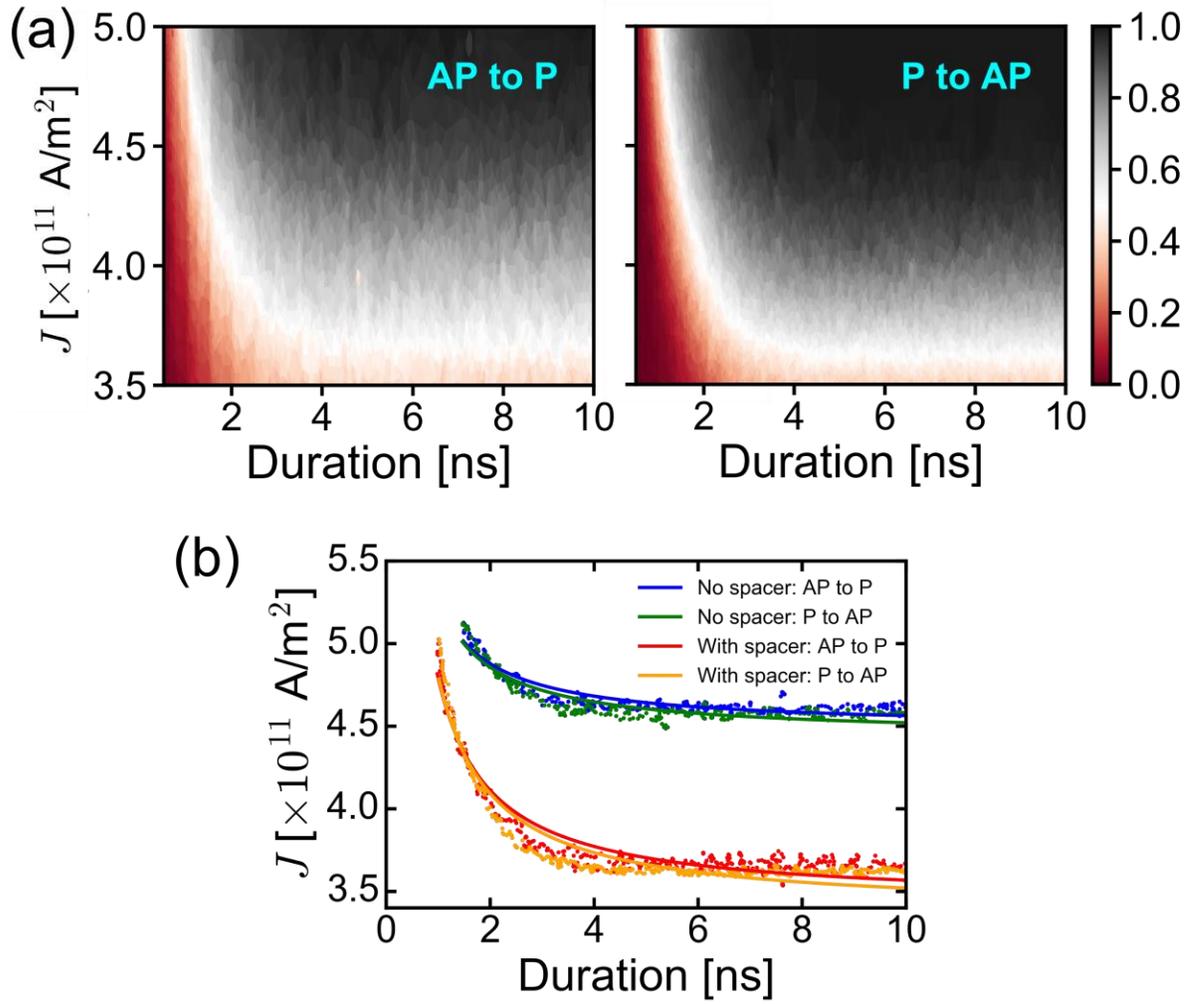

**Figure 2:** Pulsed switching measurements on PtHf 3T-MTJ devices. (a) Switching phase diagrams of a device with the 0.7 nm Hf spacer for (upper panel) AP-to-P and (lower panel) P-to-AP switching. (b) 50% switching probability points for the devices with and without a Hf spacer. The solid lines show fits to the macrospin model. The fit parameters are summarized in Table 1.

**Table 1:** Fit parameters from ferromagnetic resonance measurements, and DC and pulsed switching measurements of PtHf 3T-MTJ devices, in comparison with the Pt device reported in Ref. [10]. All devices have the same MTJ sizes of 45 nm × 190 nm (aspect ratio = 1:4.2).

| Structure | Pt(5) \| Hf(0.7) \| FCB(1.6) Pt with spacer[10] | PtHf(6) \| FCB(1.4) PtHf no spacer | PtHf(6) \| Hf(0.7) \| FCB(1.4) PtHf with spacer |
|---|---|---|---|
| *FMR measurements* | | | |
| $M_{eff}$ [Oe] | 4165 | 3340 | 3620 |
| $\alpha$ [×10$^{-3}$] | 18 | 26 ± 2 | 17 ± 2 |
| *DC switching measurements* | | | |
| $\Delta$ [$k_B T$] | 54 ± 5 | 70 ± 13 | 36 ± 8 |
| $J_0^{DC}$ [×10$^{11}$A/m$^2$] | 4.0 | 2.2 ± 0.2 | 1.4 ± 0.1 |
| $\xi_{DL}$ | 0.055 | 0.098 | 0.106 |
| *Pulse switching measurements* | | | |
| $J_0^{pulse}$ [×10$^{11}$A/m$^2$] | 6.3 | 4.5 | 3.4 |
| $t_0$ [ns] | 1.1 | 0.2 | 0.4 |

# SUPPLEMENTAL INFORMATION

# Efficient Switching of 3-Terminal Magnetic Tunnel Junctions by the Giant Spin Hall Effect of Pt$_{85}$Hf$_{15}$ Alloy

Minh-Hai Nguyen[1], Shengjie Shi[1], Graham E. Rowlands[2], Sriharsha V. Aradhya[1*],

Colin L. Jermain[1], D. C. Ralph[1,3], R. A. Buhrman[1†]

[1] *Cornell University, Ithaca, New York 14853, USA*

[2] *Raytheon BBN Technologies, Cambridge, Massachusetts 02138, USA*

[3] *Kavli Institute at Cornell, Ithaca, New York 14853, USA*

[*]Currently at Western Digital Corp., Fremont, CA 94539
[†] Email address: buhrman@cornell.edu

**S1. Ferromagnetic resonance measurements**

To estimate the Gilbert magnetic damping and effective demagnetization field of our samples, we performed flip-chip ferromagnetic resonance (FMR) measurements on unpatterned portions of our multilayer stacks. The technique is fully described in the Supplemental Information of Ref. [1]. Figure S1(a) shows a typical FMR lineshape for the PtHf sample with the Hf spacer at the excitation frequency $f = 10$ GHz. By fitting to the derivative Lorentzian function, the resonance field $H_0$ and linewidth $LW$ of the FMR lineshape were obtained and plotted in Fig. S1(b,c). The effective field $M_{eff}$ was estimated by fitting to the Kittel equation:

$$f = \frac{\gamma}{2\pi} \mu_0 \sqrt{H_0(H_0 + M_{eff})}, \qquad (S1)$$

where $\gamma$ is the gyromagnetic ratio (also a fitting parameter) and $\mu_0 = 4\pi \times 10^{-7}$ N/A$^2$ the vacuum permeability. The damping coefficient $\alpha$ is similarly estimated by a linear fit to the relation:

$$LW = f \cdot \frac{2\pi}{\gamma} \alpha + LW_0. \qquad (S2)$$

The results are listed in Table 1 in the main text.

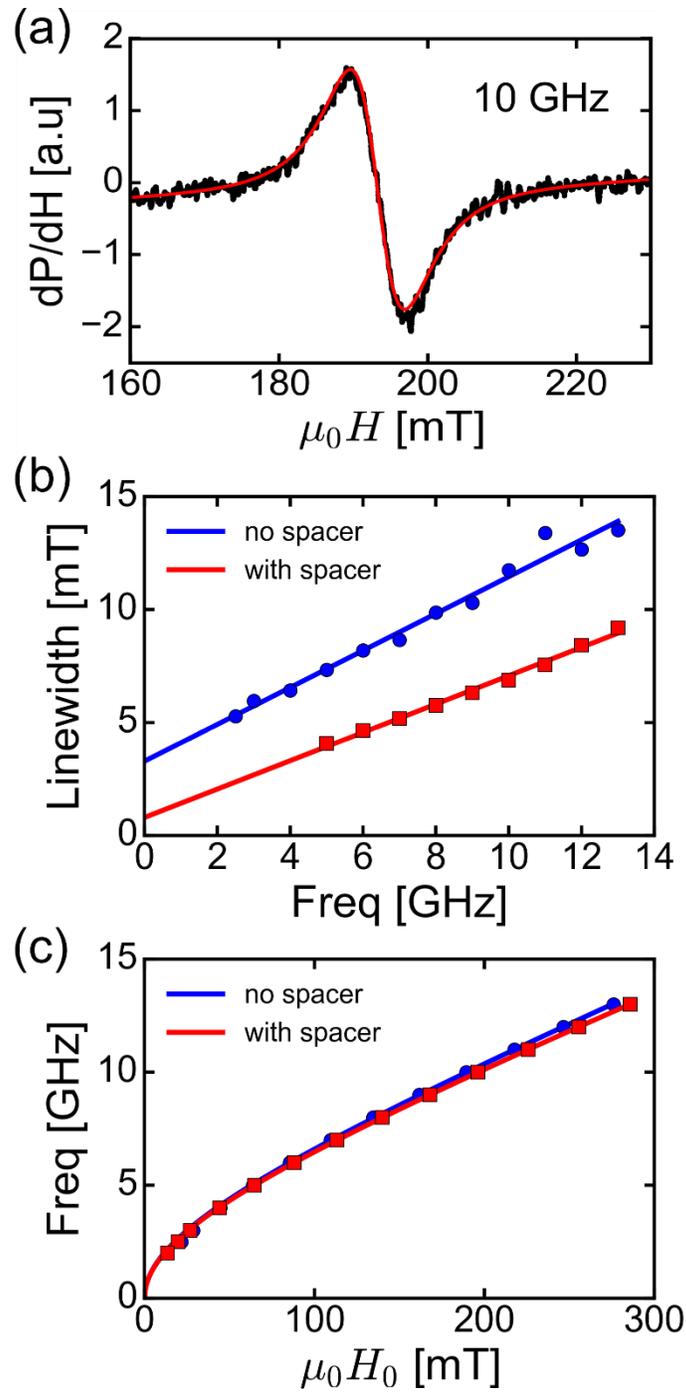

**Figure S1:** (a) A typical FMR lineshape (shown for a sample with the 0.7 nm Hf spacer at 10 GHz) and a fitted line to the derivative Lorentzian function. (b) Linewidth of the FMR signal as a function of frequency. Solid lines show linear fit results. (c) Frequencies versus measured resonance fields. Solid lines show fits to the Kittel formula.

## S2. Pulsed switching measurement methods

Figure S2 shows a schematic diagram of the pulsed switching circuit used in our measurements. The state of the MTJ is determined by measuring the voltage across the tunnel barrier when a small current 1 µA is applied to flow through it by a Keithley 2450 digital multimeter. The read voltages are then stored in the internal memory of a NI-DAQmx USB-6361 (up to 64MB) until transferred to the controlling computer through an USB connection. The NI-DAQ is also responsible for sending trigger pulses to two Picosecond Pulse Labs generators for generating reset (PSPL 10070A) and switching (PSPL 10100A) pulses. The switching pulses from the PSPL with resolution of 1 dB go through a Mini-Circuits voltage controlled attenuator (VCA ZX73), which is supplied and controlled by the DAC output of a Signal Recovery 7265 DSP Lock-in amplifier and a Yokogawa 7651 power supply, before being combined with the reset pulses using a Picosecond 5331 power splitter. A GMW 5403 Electromagnet is used to apply an external field parallel to the easy axis of the MTJ.

For each switching attempt, a reset pulse (of opposite sign to the switching pulse) is sent to the channel to reset the state of the MTJ before a switching pulse. The switching probability is determined by the ratio of successfully switched counts over the successfully reset counts in 1000 attempts. The switching probability points are grouped into non-overlapping triangles by Delaunay triangulation, each of which is colored based on the average value of its vertices. We also employed the adaptive measurement strategy which is described in detail in section S3.

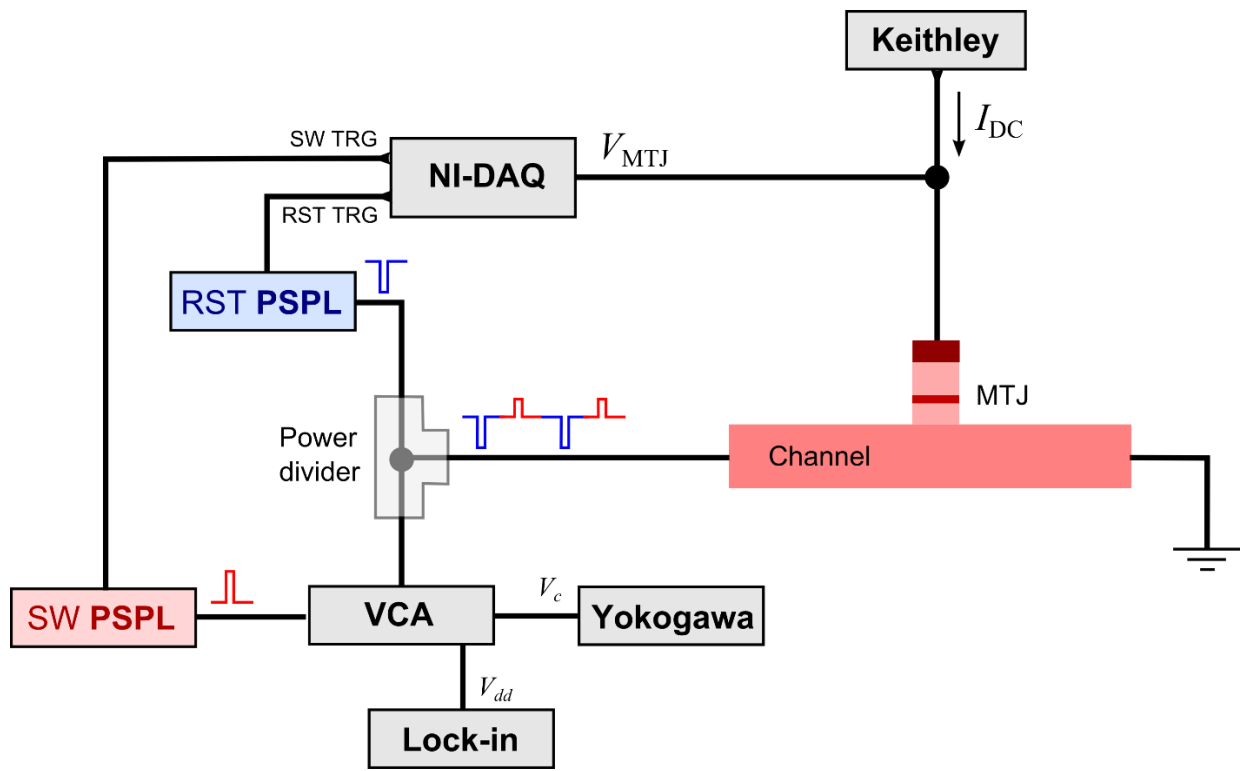

**Figure S2:** Diagram of the pulsed switching measurement circuit.

## S3. Adaptive measuring method

To most efficiently measure the switching phase diagrams for our devices, we deployed an adaptive measurement method which is illustrated by the flow chart in Fig. S3. We note that in the switching phase diagrams as shown in Fig. 2(a) (main text), the richest information is located in the vicinity of the 50% switching probability boundary between the non-switched (red) at the lower left corner and nearly 100% switched (black) at the upper right corner. Thus, it is maximally time-effective if most measuring points lie along the 50% probability boundary, instead of being distributed evenly over the *V-t* space. To achieve this, we carry out the experiment in multiple iterations (repetitions) after each of which the data set is collected, analyzed for determining the parameters for the next one. The effectiveness of this method is to tailor the strategy for setting parameters so that the data points are gathered in the areas (of the parameter space) of richest information.

Assume after one iteration we obtain a series of probability $P_i(V_i,t_i)$. To visually plot the data, we group the data points *($V_i,t_i$)* into triangles by Delaunay triangulation, each of which is colored based on the average value of its vortices, as illustrated in Fig. S4(b). We define a "weight function" $f_{i,j}=f(P_i,P_j)$ for each edge *($V_i,t_i$)* and *($V_j,t_j$)*. For the edges whose weight meets a pre-defined criterion (in our case, the top 33% of all edges), we add their midpoints to the list of data points to be measured in the next iterations. The finishing condition is set to be the maximum number of iterations. Fig. S4(b-d) shows the measuring progress during which the newly added points are located near the 50% probability boundary. Fig. S4(e) shows the final result (without illustrative lines).

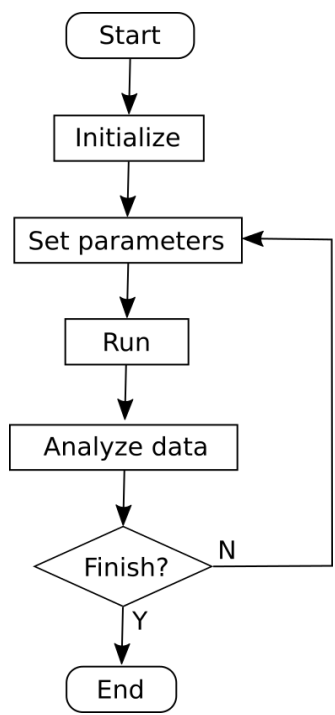

**Figure S3:** Flow chart of the adaptive measuring automation program. The experimental data set after each run (iteration) with a given set of parameters is analyzed for determining the parameters for the next iteration (if necessary). This protocol focuses measurement resources efficiently on the areas in parameters space with the most information content.

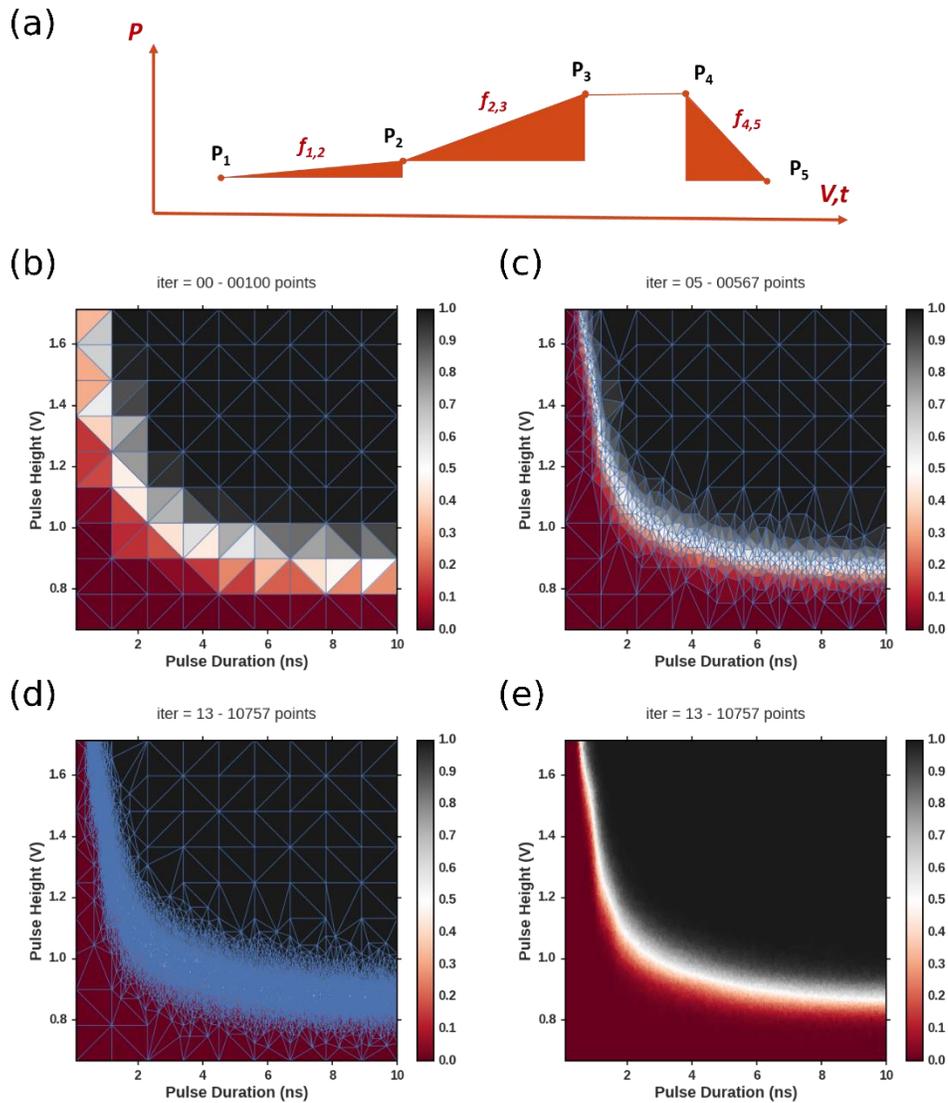

**Figure S4:** Example of the adaptive method in measuring switching phase diagrams. (a) The weight function is determined by the area of the "triangle" (integral) made by the switching probability and the 2 parametric axes. (b) The initial distribution of data points, which is uniform over the *V-t* space. The points are grouped into non-overlapping triangles (thin lines) by Delaunay triangulation, each of which is colored based on the average value of its 3 vertices. (c) Results after 5 iterations (567 points). (d) Results after 13 iterations (10757 points) in which most data points lie along the 50% probability boundary. (e) Results as in (d) with the illustrative lines removed.